%% file: paper.tex
\documentclass[runningheads]{llncs}
\usepackage{graphicx}
\usepackage{cite}

\usepackage{booktabs}
\usepackage{multirow}
\usepackage{makecell}
\usepackage{tabularx}
\usepackage{subfig}

\begin{document}
\title{Brain Tumor Segmentation using 3D-CNNs with Uncertainty Estimation}

\author{Laura Mora Ballestar \and Veronica Vilaplana
\thanks{This work has been partially supported by the project MALEGRA TEC2016-75976-R financed by the Spanish Ministerio de Econom\'{i}a y Competitividad.} }

\authorrunning{L. Mora et al.}
%
\institute{Signal Theory and Communications Department, Universitat Politècnica de Catalunya. BarcelonaTech, Spain \\
\email{lmoraballestar@gmail.com, veronica.vilaplana@upc.edu}\\
}

\maketitle              
\begin{abstract}
\textit{Automation of brain tumors in 3D magnetic resonance images (MRIs) is key to assess the diagnostic and treatment of the disease. In recent years, convolutional neural networks (CNNs) have shown improved results in the task. However, high memory consumption is still a problem in 3D-CNNs. Moreover, most methods do not include uncertainty information, which is specially critical in medical diagnosis. This work proposes a 3D encoder-decoder architecture, based on V-Net \cite{vnet} which is trained with patching techniques to reduce memory consumption and decrease the effect of unbalanced data. We also introduce voxel-wise uncertainty, both epistemic and aleatoric using test-time dropout and data-augmentation respectively. Uncertainty maps can provide extra information to expert neurologists, useful for detecting when the model is not confident on the provided segmentation.}

\end{abstract}

\input{sections/introduction.tex}
\input{sections/sota.tex}
\input{sections/method.tex}
\input{sections/results.tex}
\input{sections/conclusions}

%
%
%

\end{document}

%% file: sections/introduction.tex
\section{Introduction}

Brain tumors are categorized into primary, brain originated; and secondary, tumors that have spread from elsewhere and are known as brain metastasis tumors. Among malignant primary tumors, gliomas are the most common in adults, representing 81\% of brain tumors \cite{epidemiology}. The World Health Organization (WHO) categorizes gliomas into grades I-IV which can be simplified into two types (1) “low grade gliomas” (LGG), grades I-II, which are less common and are characterized by low blood concentration and slow growth and (2) “high grade gliomas” (HGG), grades III-IV, which have a faster growth rate and aggressiveness.

The extend of the disease is composed of four heterogeneous histological sub-regions, i.e. the peritumoral edematous/invaded tissue, the necrotic core (fluid-filled), the enhancing and no-enhancing tumor (solid) core. Each region is described by varying intensity profiles across MRI modalities (T1-weighted, post-contrast T1-weighted, T2-weighted, and Fluid-Attenuated Inversion Recovery-FLAIR), which reflect the diverse tumor biological properties and are commonly used to assess the diagnosis, treatment and evaluation of the disease. These MRI modalities facilitate tumor analysis, but at the expense of performing manual delineation of the tumor regions which is a challenging and time-consuming process. For this reason, automatic mechanisms for region tumor segmentation have appeared in the last decade thanks to the advancement of deep learning models in computer vision tasks.

The Brain Tumor Segmentation (BraTS) \cite{brats, advancing_brats, bakas_identifying, segmentation_label_gbm, segmentation_label_lgg} challenge started in 2012 with a focus on evaluating state-of-the-art methods for glioma segmentation in multi-modal MRI scans. BraTS 2020 training dataset includes 369 cases (293 HGG and 76 LGG), each with four 3D MRI modalities rigidly aligned, re-sampled to $1 mm^3$ isotropic resolution and skull-stripped with size $240x240x155$. Each provides a manual segmentation approved by experienced neuro-radiologists. Training annotations comprise the enhancing tumor (ET, label 4), the peritumoral edema (ED, label 2), and the necrotic and non-enhancing tumor core (NCR/NET, label 1). The nested sub-regions considered for evaluation are: whole tumor WT (label 1, 2, 4), tumor core TC (label 1, 4) and enhancing tumor ET (label 4). The validation set includes 125 cases, with unknown grade nor ground truth annotation.

This work describes a semantic segmentation approach for 3D brain tumor segmentation and uncertainty estimation. We use the V-Net \cite{vnet} architecture to segment the three sub-regions and we estimate both epistemic and aleatory \cite{unc_categorization} uncertainties using test-time dropout (TTD)\cite{montecarlo_dropout} and data augmentations, respectively.

%% file: sections/sota.tex
\section{Related Work}

\subsection{Semantic Segmentation}
Brain tumor segmentation methods include generative and discriminative approaches. Generative methods try to incorporate prior knowledge and model probabilistic distributions whereas discriminative methods extract features from image representations. This latter approach has thrived in recent years thanks to the advancement in CNNs, as demonstrated in the winners of the previous BraTS. The biggest break through in this area was introduced by DeepMedic \cite{deepmedic} a 3D CNN that exploits multi-scale features using parallel pathways and incorporates a fully connected conditional random field (CRF) to remove false positives. \cite{Casamitjana1} compares the performances of three 3D CNN architectures showing the importance of the multi-resolution connections to obtain fine details in the segmentation of tumor sub-regions.
More recently, EMMA \cite{ensembles} creates and ensemble at inference time which reduces overfitting but at high computational cost, and \cite{Casamitjana2} proposes a cascade of two CNNs, where the first network produces raw tumor masks and the second network is trained on the vecinity of the tumor to predict tumor regions. BraTS 2018 winner \cite{myronenko20183d} proposed an asymmetrically encoder-decoder architecture with a variational autoencoder to reconstruct the image during training, which is used as a regularizer. Isensee, F \cite{no_newnet} uses a regular 3D-U-Net optimized on the evaluation metrics and co-trained with external data. BraTS 2019 winners \cite{two-stage} use a two-stage cascade U-Net trained end-to-end. Finally, \cite{tricks} applies several tricks in three categories: data processing, model devising and optimization process to boost the model performance.

\subsection{Uncertainty}

Uncertainty information of segmentation results is important, specially in medical imaging, to guide the clinical decisions and help understand the reliability of the provided segmentation, hence being able to identify more challenging cases which may require expert review. Segmentation models for brain tumor MRIs tend to label voxels with less confidence in the surrounding tissues of the segmentation targets \cite{wangUnc}, thus indicating regions that may have been miss-segmented.

Last year's BraTS challenge already started introducing uncertainty measurements. \cite{demistifying} computes epistemic uncertainty using TTD. They obtain a posterior distribution generated after running several epochs for each image at test-time. Then, mean and variance are used to evaluate the model uncertainty. A different approach is proposed by Wang G \cite{wangUnc}, who uses TTD and data augmentation to estimate the voxel-wise uncertainty by computing the entropy instead of the variance. Finally, \cite{McKinleyUnc} proposes to incorporate uncertainty measures during training as they define a loss function that models label noise and uncertainty.

%% file: sections/method.tex
\section{Method}
The following section details the network architecture as well as the training schemes and data processing techniques used to train our model. 

\subsection{Data Pre-processing and Augmentation}
MRI intensity values are not standardized as the data is obtained from different institutions, scanners and protocols. When training a neural network it is particularly important to use normalized data, even more as MRI modalities are treated like color channels. For this reason, we normalize each modality of each patient independently to have zero mean and unit std based on non-zero voxels only, which represent the brain region.

We also apply data augmentation techniques to prevent over-fitting by trying to disrupt minimally the data. For this, we apply Random Flip (for all 3 axes) with a 50\% probability, Random Intensity Shift between ($-0.1..0.1$ of data std) and Random Intensity Scale on all input channels at range (0.9..1.1).

\subsection{Network Architecture}
V-Net \cite{vnet} and U-Net \cite{unet} architectures have proven to be successful and reliable encoder-decoder architectures for medical image segmentation, as they are able to segment fine image structures. Moreover, may past years participants achieve great results using this architectures as baselines. With this in mind, our work uses a V-Net architecture with four output channels and some minor modifications, such as the usage of Instance Normalization \cite{ulyanov2016instance} in contrast of Batch Normalization, which normalizes across each channel for each training example. Moreover we have doubled the number of features maps at each level of the network as proposed in \cite{no_newnet}, having 32 feature channels at the highest resolution.

\begin{figure}[ht]
    \centering
    \includegraphics[width=1\textwidth]{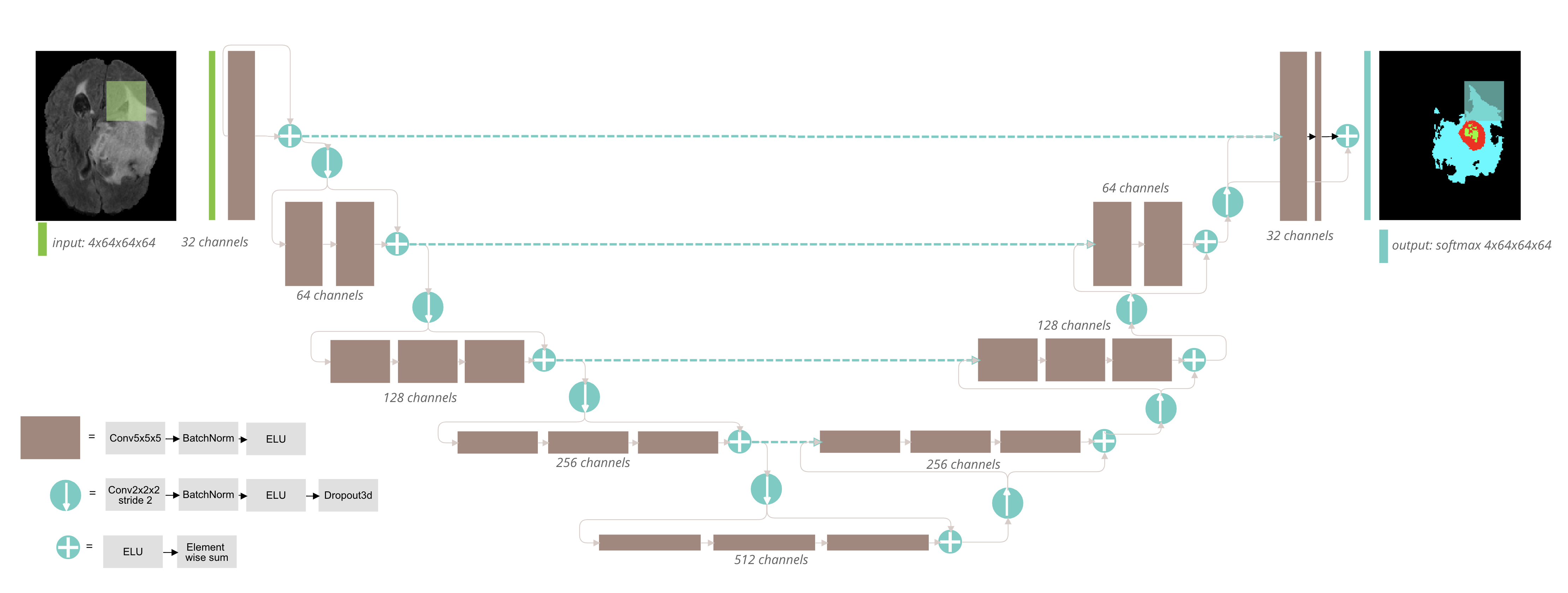}
    \caption{We use the V-Net \cite{vnet} architecture with instance normalization, ELU non-linearities and doubled number of feature channels. Feature dimensionality is denoted at each block. The network outputs the segmentation in four channels using a Softmax.} 
    \label{fig:network}
\end{figure}

The network follows a common approach that progressively downsizes the image and feature dimensions by a factor of 2 using strided convolutions instead of pooling layers, see Figure \ref{fig:network}.

\subsection{Training}

The network is trained end-to-end with randomly selected patches which have a 50\% probability being centered on healthy tissue and 50\% probability on tumor \cite{efficient_kamnitsas}. Due to memory constraints, we need to make a trade-off between the size of each patch and the batch size. With this, we found that the best results are achieved with patches of size 64x64x64 and a batch size of 8. In our case, bigger patch sizes required smaller batches and were more likely to overfit, thus achieving worst results on the validation set.

As optimization, we use SGD with $0.99$ momentum and a learning rate of $1e-2$. We use the Pytorch learning scheduler ReduceLROnPlateau, which will decrease by a factor of $0.1$ whenever the validation loss has not improved in the passed 10 epochs.

For the loss function we use the dice loss as defined in \cite{vnet} which can be written as

\[ L_{dice} =  \frac{2*\sum_{i}^{N}p_{i}g_{i}}{\sum_{i}^{N}p_{i}^{2}+\sum_{i}^{N}g_{i}^{2} + \epsilon } \]
where  N is the number of voxels, $p_{i}$  and $g_{i}$ correspond to the predicted and ground truth labels per voxel respectively, and $\epsilon$ is added to avoid zero division.

We also train a version of the model where the loss is optimized on the three nested regions, whole tumor, tumor core and enhancing tumor together with the previous dice loss.

\subsection{Post-Processing}
The proposed model shows several false positives in the form of small and separated connected components. Therefore,
we keep the two biggest connected components if their proportion is bigger than some threshold (obtained by analysing the training set), as some of the subjects may have several regions with unhealthy tissue. 


\subsection{Uncertainty}
This year's BraTS includes a third task to evaluate the model uncertainty and reward methods with predictions that are: (a) confident when correct and (b) uncertain when incorrect. In this work, we model the voxel-wise uncertainty of our method at test time, using test time dropout (TTD) and test-time data augmentation for epistemic and aleatoric uncertainty respectively.

We compute epistemic uncertainty as proposed in Gal et.al \cite{gal2015dropout}, who uses dropout as a Bayesian Approximation in order to simplify the task. Therefore, the idea is to use dropout both at training and testing time. The paper suggests to repeat the prediction a few hundred times with random dropout. Then, the final prediction is the average of all estimations and the uncertainty is modelled by computing the variance of the predictions. In this work we perform $B=50$ iterations and use dropout with a 50\% probability to zero out a channel. As said, the uncertainty map is estimated with the variance for each sub-region independently. Let $Y^{i} = \left \{ y_{1}^{i}, y_{2}^{i}...y_{B}^{i} \right \}$ be the vector that represents the i-th voxel's predicted labels, the voxel-wise uncertainty map, for each evaluation region, is obtained as the variance:

\[var = \frac{1}{B} \sum_{b=1}^{B} (y_{b}^{i} - y_{mean}^{i})^{2}\]

Uncertainty can also be estimated with the entropy, as \cite{wangUnc} showed. However, the entropy will provide a global measure instead of map for sub-region. In this case, the voxel-wise uncertainty is calculated as:

\[H(Y^{i}|X) \approx - \sum_{m=1}^{M} \hat{p}_{m}^{i}\ln(\hat{p}_{m}^{i})\]

where $\hat{p}_{m}^{i}$ is the frequency of the m-th unique value in $Y^{i}$ and $X$ represents the input image.

Finally, we model aleatoric uncertainty by applying the same augmentation techniques from training plus random Gaussian noise. The final prediction and uncertainty maps are computed following the same strategy as in the epistemic uncertainty.

%% file: sections/results.tex
\section{Results}

The model has been implemented in Pytorch \cite{torch} and trained on the GPI\footnote{The UPC Image and Video Processing Group (GPI) is a research group of the Signal Theory and Communications department.} servers, based on 2 Intel(R) Xeon(R) @ 2.40GHz CPUs using 16GB RAM and a 12GB NVIDIA GPU, using BraTS 2020 training dataset. We report results on both training (369 cases) and validation (125 cases) datasets. All results, prediction and uncertainty maps, are uploaded to the CBICA’s Image Processing Portal (IPP)\footnote{BraTS20 leaderboard: https://www.cbica.upenn.edu/BraTS20/} for evaluation of Dice score, Hausdorff distance (95th percentile), sensitivity and specificity per each class. Specific uncertainty evaluation metrics are the ratio of filtered TN (FTN) and the ratio of filtered TP. 

\subsection{Segmentation}

The principal metrics to evaluate the segmentation performance are the Dice Score, which is an overlap measure for pairwise comparison of segmentation mask X and ground truth Y:

\[  DSC = 2* \frac{\left | X \cap  Y \right |}{|X| + |Y|} \]

and the Hausdorff distance, which is the maximum distance of a set to the nearest point in the other set, defined as:

\[D_{H}(X,Y) = max\left \{ sup_{x\epsilon X} \inf _{y\epsilon Y} d(x,y)) , sup_{y\epsilon Y} \inf _{x\epsilon X} d(x,y)) \right \} \]

where sup represents the supremum and inf the infimum. In order to have more robust results and to avoid issues with noisy segmentation, the evaluation scheme uses the 95th percentile.

Table \ref{tab:train} and Table \ref{tab:validation} show results for training and validation sets respectively. We show the segmentation maps obtain directly from our model and the ones obtained from averaging the predictions from uncertainty estimation, annotated with (avg).

\begin{table}[ht]
    \caption{Segmentation Results on Training Dataset (369 cases).}\label{tab:train}
    \centering
    \begin{tabular}{l|c|c|c|c|c|c}
        \hline
        \multirow{2}{*}{\textbf{Method}} & \multicolumn{3}{l|}{\textbf{Dice}} & \multicolumn{3}{l}{\textbf{Hausdorff (mm)}} \\[1ex]
        \cline{2-7}
         & WT & TC & ET & WT & TC & ET \\
        \hline\hline
        V-Net           &	0.8421	      & 0.7837           & 0.6752	     &	21.9354         &	11.7565 & 31.1815 \\ [1ex]
        V-Net+post      & \textbf{0.8513} & 0.7852	        & \textbf{0.6765}&	\textbf{16.8425} &	\textbf{10.8923} &	\textbf{30.6484} \\[1ex]
        V-Net (avg)     & 0.8414	      & 0.7727          & 0.6482        & 22.7816          & 11.8985    & 34.0967  \\ [1ex]
        V-Net+post (avg)& 0.8342          & \textbf{0.7860}	& 0.6635	     & 19.2272 	       & 11.3911    & 34.2888 	 \\[1ex]
        
        \hline
    \end{tabular}
\end{table}

\begin{table}[ht]
    \caption{Segmentation Results on Validation Dataset (125 cases)}\label{tab:validation}
    \centering
    \begin{tabular}{l|c|c|c|c|c|c}
        \hline
        \multirow{2}{*}{\textbf{Method}} & \multicolumn{3}{l|}{\textbf{Dice}} & \multicolumn{3}{l}{\textbf{Hausdorff (mm)}} \\[1ex]
        \cline{2-7}
         & WT & TC & ET & WT & TC & ET \\
        \hline\hline
        V-Net &  0.8368 &	0.7499	     &  0.6159 &	26.4085	& 13.3398 & 49.7425 \\ [1ex]
        V-Net+post & \textbf{0.8463} &	\textbf{0.7526}   & 0.6179 &	20.4073 & \textbf{12.1752} & \textbf{47.7020} \\ [1ex]
        V-Net (avg) & 0.8335	& 0.7547	& \textbf{0.6215} & 26.1902 & 13.1222 & 50.1156 \\ [1ex]
        V-Net+post (avg) & 0.8421 & 0.7503	& 0.6175	& 2\textbf{0.4005} & 12.9294 & 48.7698  \\ [1ex] 
        \hline
    \end{tabular}
\end{table}

In both sets, the best results are obtained with the V-Net model directly when post-processing is applied. However, both Dice score and Hausdorff distance have lower performance on the validation set, being more noticeable in the ET region, were it increments from 31 to 47. This high distance is caused because we obtain 13 predictions with the highest Hausdorff distance and 0 dice, indicating we are predicting ET in tumors that should not have it.

\subsection{Uncertainty}

BraTS requires to upload three uncertainty maps, one for each subregion (WT, TC, ET) together with the prediction map. Moreover, uncertainty maps must be normalized from 0-100 such that "0" represents the most certain prediction and "100" represents the most uncertain. Moreover they measure the FTP defined as $FTP = (TP_{100} - TP_{T}) / TP_{100}$, where T represents the threshold used to filter the more uncertain values. The ratio of filtered true negatives (FTN) is calculated in a similar manner.

\begin{table}[ht]
    \caption{Uncertainty Results on Training and Validation Dataset}\label{tab:validation_unc}
    \centering
    \begin{tabular}{l|c|c|c||c|c|c||c|c|c}
        \hline
        \multirow{2}{*}{\textbf{Method}} & \multicolumn{3}{c||}{\textbf{DICE AUC}} & \multicolumn{3}{c||}{\textbf{FTP RATIO AUC}} & \multicolumn{3}{c}{\textbf{FTN RATIO AUC}} \\[1ex]
         \cline{2-10}
         & WT & TC & ET & WT & TC & ET & WT & TC & ET \\
        \hline\hline
        (train) V-Net+post & 0.8506 & 0.7890 & 0.6779 & 0.0029 & 0.0104 & 0.0179 & 0.0008 & 0.0002 & 0.0002 \\ [1ex]
        (valid) V-Net+post & 0.8505 & 0.7583 & 0.6274 & 0.0118 & 0.0515 & 0.0815 & 0.0027 & 0.0008 & 0.0005 \\ [1ex]
        \hline
        
    \end{tabular}
\end{table}

Results on Table \ref{tab:validation_unc} show that our model has low FTP and FTN, meaning that it is certain of those predictions.

\subsection{Visual Analysis}

\begin{figure}[t]
    \includegraphics[width=1\textwidth]{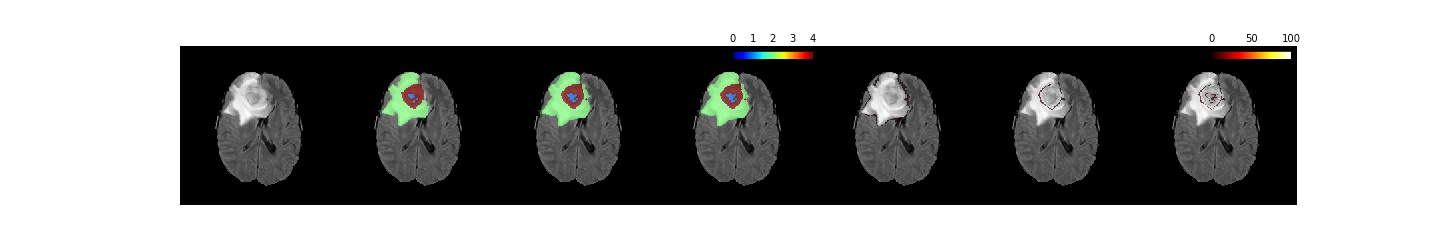}
    \includegraphics[width=1\textwidth]{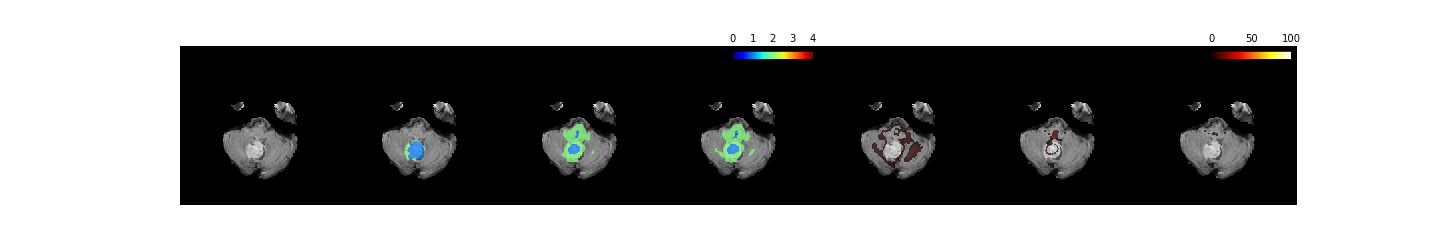}
    \caption{Training results on patients: 223 and 325 (top-bottom). Image order: (1) Flair (2) GT (3) Prediction (4) Average Prediction from uncertainty (5) WT uncertainty map (6) TC uncertainty map (7) ET uncertainty map}
    \label{fig:train_res}
    
    \includegraphics[width=1\textwidth]{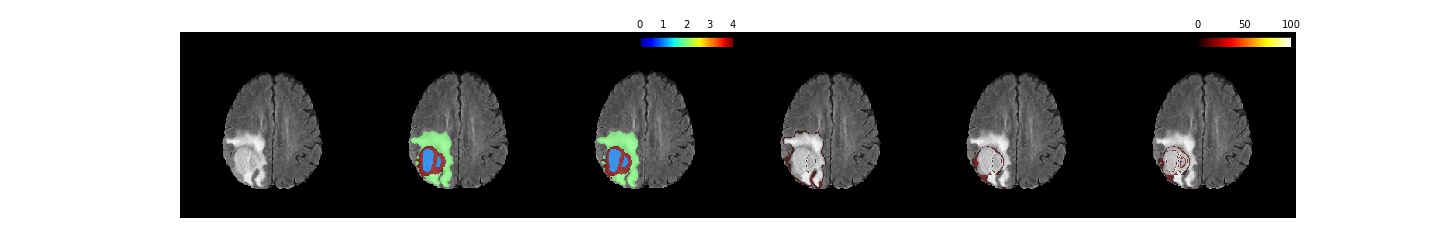}
    \includegraphics[width=1\textwidth]{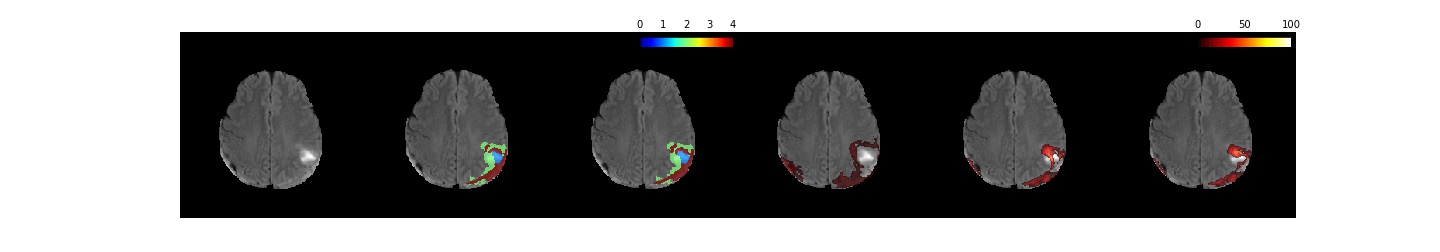}
    \caption{Validation results on patients: 007 and 035 (top-bottom). Image order: (1) Flair (2) Prediction (3) Average Prediction from uncertainty (4) WT uncertainty map (5) TC uncertainty map (5) ET uncertainty map}
    \label{fig:val_res}
\end{figure}

Figure \ref{fig:train_res} and figure \ref{fig:val_res} show some visual results from the training and validation sets respectively. As it can be seen, in some of the subjects we obtain a fairly good segmentation, whereas in others we have too many false positives in the WT and ET regions. However, the uncertainty values show that the model is more uncertain in the tumor surroundings and areas where the prediction has been miss-classified.

%% file: sections/conclusions.tex
\section{Discussion and Conclusions}

In this paper we use a V-Net architecture with minor modifications: number of feature maps and instance normalization instead of using batch normalization. Results, on both training and validation sets, prove that our model has severe problems to correctly segment whole tumor and, specially enchancing tumor. In both cases, we have a large number of false positives, as can be seen in the Haurdorff distance results. In the case of ET, the distance is much bigger as a single false positive voxel in a patient where no enhancing tumor is present in the ground truth results in a Dice score of 0 and a Hausdorff distance of approximately 373, which is the case for some of the subjects which has a dramatically effect on the mean value. 

In order to improve the results, we plan on adding ResNet blocks to the network baseline as well as using the Generalised Dice Loss \cite{gen_dice}, more suited for unbalanced data. Moreover, we believe that some post-processing techniques should also improve the results.

%% file: paper.bbl
\begin{thebibliography}{8}
\bibliographystyle{splncs04}


\bibitem{brats}
B. H. Menze, A. Jakab, S. Bauer, J. Kalpathy-Cramer, K. Farahani, J. Kirby, et al.: "The Multimodal Brain Tumor Image Segmentation Benchmark (BRATS)", IEEE Transactions on Medical Imaging 34(10), 1993-2024 (2015) \doi{ 10.1109/TMI.2014.2377694}

\bibitem{advancing_brats}
S. Bakas, H. Akbari, A. Sotiras, M. Bilello, M. Rozycki, J.S. Kirby, et al.: "Advancing The Cancer Genome Atlas glioma MRI collections with expert segmentation labels and radiomic features", Nature Scientific Data, 4:170117 (2017) \doi{10.1038/sdata.2017.117}

\bibitem{bakas_identifying}
S. Bakas, M. Reyes, A. Jakab, S. Bauer, M. Rempfler, A. Crimi, et al.: "Identifying the Best Machine Learning Algorithms for Brain Tumor Segmentation, Progression Assessment, and Overall Survival Prediction in the BRATS Challenge", arXiv preprint arXiv:1811.02629 (2018)

\bibitem{segmentation_label_gbm}
S. Bakas, H. Akbari, A. Sotiras, M. Bilello, M. Rozycki, J. Kirby, et al., "Segmentation Labels and Radiomic Features for the Pre-operative Scans of the TCGA-GBM collection", The Cancer Imaging Archive, 2017. DOI: 10.7937/K9/TCIA.2017.KLXWJJ1Q

\bibitem{segmentation_label_lgg}
S. Bakas, H. Akbari, A. Sotiras, M. Bilello, M. Rozycki, J. Kirby, et al., "Segmentation Labels and Radiomic Features for the Pre-operative Scans of the TCGA-LGG collection", The Cancer Imaging Archive, 2017. DOI: 10.7937/K9/TCIA.2017.GJQ7R0EF

\bibitem{vnet}
Milletari, Fausto, Nassir Navab, and Seyed-Ahmad Ahmadi. ”V-net: Fully convo- lutional neural networks for volumetric medical image segmentation.” 3D Vision (3DV), 2016 Fourth International Conference on. IEEE, 2016.

\bibitem{epidemiology}
Morgan, L Lloyd: The epidemiology of glioma in adults: A "state of the science" review. Neuro-oncology vol.17 01-2015 \doi{10.1093/neuonc/nou358}

\bibitem{unc_categorization}
Armen Der Kiureghian and Ove Ditlevsen: Aleatory or epistemic? does it matter? Structural safety, 31(2):105–112, 2009.

\bibitem{montecarlo_dropout}
Yarin Gal and Zoubin Ghahramani: Dropout as a bayesian approximation: Representing model uncertainty in deep learning. arXiv preprint arXiv:1506.02142, 2015


\bibitem{deepmedic}
Konstantinos Kamnitsas, Christian Ledig, Virginia F.J. Newcombe, Joanna P. Simpson, Andrew D. Kane, David K. Menon, Daniel Rueckert, Ben Glocker:
Efficient multi-scale 3D CNN with fully connected CRF for accurate brain lesion segmentation, Medical Image Analysis, Volume 36, 2017, pages 61-78,
\doi{10.1016/j.media.2016.10.004}

\bibitem{Casamitjana1} Casamitjana, A., Puch, S., Aduriz, A., Vilaplana, V., "3D Convolutional Neural Networks for Brain Tumor Segmentation: a comparison of multi-resolution architectures". In: Brainlesion: Glioma, Multiple Sclerosis, Stroke and Traumatic Brain Injuries. BrainLes 2016. Lecture Notes in Computer Science, vol 10154. Springer, 2017.

\bibitem{ensembles}
Kamnitsas, K., Bai, W., Ferrante, E., McDonagh, S., Sinclair, M., Pawlowski, N: Ensembles of Multiple Models and Architectures for Robust Brain Tumour Segmentation in International MICCAI Brainlesion Workshop (Quebec, QC), 450–462 arXiv preprint arXiv:1711.01468, 2017

\bibitem{Casamitjana2} Casamitjana, A., Catà, M., Sánchez, I., Combalia, M., Vilaplana, V., "Cascaded V-Net Using ROI Masks for Brain Tumor Segmentation". In: Brainlesion: Glioma, Multiple Sclerosis, Stroke and Traumatic Brain Injuries. BrainLes 2017. Lecture Notes in Computer Science, vol 10670. Springer, 2018.

\bibitem{myronenko20183d}
Andriy Myronenko: 3D MRI brain tumor segmentation using autoencoder regularization. arXiv preprint arXiv:1810.11654, 2016

\bibitem{no_newnet}
Isensee, F., et al.: No new-net. International MICCAI Brainlesion Workshop, pp. 234–244. Springer (2018)

\bibitem{two-stage}
Jiang Z., Ding C., Liu M., Tao D. (2020) Two-Stage Cascaded U-Net: 1st Place Solution to BraTS Challenge 2019 Segmentation Task. In: Crimi A., Bakas S. (eds) Brainlesion: Glioma, Multiple Sclerosis, Stroke and Traumatic Brain Injuries. BrainLes 2019. Lecture Notes in Computer Science, vol 11992. Springer, Cham

\bibitem{tricks}
Zhao YX., Zhang YM., Liu CL. (2020) Bag of Tricks for 3D MRI Brain Tumor Segmentation. In: Crimi A., Bakas S. (eds) Brainlesion: Glioma, Multiple Sclerosis, Stroke and Traumatic Brain Injuries. BrainLes 2019. Lecture Notes in Computer Science, vol 11992. Springer, Cham

\bibitem{demistifying}
Natekar Parth, Kori Avinash, Krishnamurthi Ganapathy
AUTHOR=Natekar Parth, Kori Avinash, Krishnamurthi Ganapathy: Demystifying Brain Tumor Segmentation Networks: Interpretability and Uncertainty Analysis. Frontiers in Computational Neuroscience vol.14 page 6 \doi{10.3389/fncom.2020.00006}, 2020


\bibitem{wangUnc}
Wang G., Li W., Ourselin S. and Vercauteren T. Automatic Brain Tumor Segmentation Based on Cascaded Convolutional Neural Networks With Uncertainty Estimation. Frontiers in Computational Neuroscience vol.13 pages 56 \doi{10.3389/fncom.2019.00056}, 2019


\bibitem{McKinleyUnc}
McKinley R., Meier R., Wiest R. (2019) Ensembles of Densely-Connected CNNs with Label-Uncertainty for Brain Tumor Segmentation. In: Crimi A., Bakas S., Kuijf H., Keyvan F., Reyes M., van Walsum T. (eds) Brainlesion: Glioma, Multiple Sclerosis, Stroke and Traumatic Brain Injuries. BrainLes 2018. Lecture Notes in Computer Science, vol 11384. Springer

\bibitem{ulyanov2016instance}
Dmitry Ulyanov and Andrea Vedaldi and Victor Lempitsky. Instance Normalization: The Missing Ingredient for Fast Stylization. arXiv preprint arXiv:1607.08022, 2016

\bibitem{efficient_kamnitsas}
Kamnitsas, K., Ledig, C., Newcombe, V.F., Simpson, J.P., Kane, A.D., Menon, D.K., Rueckert, D., Glocker, B.: Efficient multi-scale 3D CNN with fully connected CRF for accurate brain lesion segmentation. Med. Image Anal. 36 (2017) 61–78

\bibitem{gal2015dropout}
Yarin Gal and Zoubin Ghahraman.Dropout as a Bayesian Approximation: Representing Model Uncertainty in Deep Learning. arXiv preprint arXiv:1506.02142, 2015

\bibitem{torch}
Paszke, Adam and Gross, Sam and Chintala, Soumith and Chanan, Gregory and Yang, Edward and DeVito, Zachary and Lin, Zeming and Desmaison, Alban and Antiga, Luca and Lerer, Adam. Automatic differentiation in PyTorch, NIPS-W 2017

\bibitem{unet}
Ronneberger, Olaf, Philipp Fischer, and Thomas Brox. ”U-net: Convolutional net- works for biomedical image segmentation.” MICCAI. Springer, 2015. \doi{10.1007/978-3-319-24574-4\_28}

\bibitem{gen_dice}
Sudre, C.H., Li, W., Vercauteren, T., Ourselin, S., Jorge Cardoso, M.Generalised Dice Overlap as a Deep Learning Loss Function for Highly Unbalanced Segmentations.  Lecture Notes in Computer Science 240-248, Springer International Publishing 2017. \doi{10.1007/978-3-319-67558-9\_28}


\end{thebibliography}
